\documentstyle[aps,pre,multicol,psfig]{revtex}

\begin{document}
\draft 

\title{Zipping and collapse of diblock copolymers}

\author{Marco Baiesi$^{(1)}$, Enrico Carlon$^{(1)}$, Enzo Orlandini$^{(1)}$ 
and Attilio L. Stella$^{(1,2)}$}
\address{
$^{(1)}$ INFM - Dipartimento di Fisica, Universit\`a di Padova, I-35131
Padova, Italy \\
$^{(2)}$ Sezione INFN, Universit\`a di Padova, I-35131 Padova, Italy
}

\date{\today}

\maketitle

\begin{abstract}
Using exact enumeration methods and Monte Carlo simulations we study the phase 
diagram relative to the conformational transitions of a two dimensional diblock 
copolymer. The polymer is made of two homogeneous strands of monomers of 
different species which are joined to each other at one end.
We find that depending on the values of the energy parameters in the model, 
there is either a first order collapse from a swollen to a compact phase of 
spiral type, or a continuous transition to an intermediate zipped phase 
followed by a first order collapse at lower temperatures. 
Critical exponents of the zipping transition are computed and their exact 
values are conjectured on the basis of a mapping onto percolation geometry,
thanks to recent results on path-crossing probabilities.
\end{abstract}
\pacs{PACS numbers: 61.41.+e, 05.70.Jk, 64.60.Ak, 64.60.Kw }

\begin{multicols}{2} \narrowtext

\section{Introduction}
\label{sec:intro}

Polymers in solution typically undergo a coil - globule transition from a high 
temperature (T) swollen phase to a low T phase where the polymer assumes 
compact conformations. In the case of homopolymers, for which all the monomers 
are identical, this transition is by now well understood \cite{degennes,carlo}.
It is known as $\Theta$-collapse and has been widely investigated in the past 
years using various methods such as mean field approximations \cite{meanfield},
exact enumerations of interacting self-avoiding walks on lattices \cite{series},
Monte Carlo \cite{MC}, transfer matrix \cite{TM} and field theoretical 
calculations \cite{fieldtheory}. 
In two dimensions the exponents of the $\Theta$-collapse have been related 
to the fractal properties of the percolation cluster and are believed to be 
known exactly \cite{percolation}.

The study of the conformational properties and phase transitions of 
macromolecules with inhomogeneous or random sequences of monomers is an 
interesting frontier in nowadays polymer statistics \cite{hetheropolymers}. 
These systems pose new theoretical and numerical challenges,
compared to their more standard, homogeneous counterparts.
Particularly interesting is in general the possibility that the 
inhomogeneities along the chain could lead to transitions and universality 
classes of scaling behavior, which are not realized for homopolymers
\cite{hethero2,monari}.
Moreover, the most complex versions of models of this class are also 
expected to be useful for the description of phenomena like protein folding 
\cite{folding}, DNA denaturation \cite{denaturation} and RNA secondary
structure formation \cite{RNA}. Thus,
an investigation of the universal properties of the simplest
among these systems can offer an important gauge of the relevant 
model ingredients necessary in order to reproduce the 
basic conformational mechanisms in more sophisticated descriptions.

One of the most elementary conformational transitions (not realized in 
homopolymers) one can try to describe in relatively simple terms is what 
we call here a {\it zipping} transition. By zipping we mean a process in which 
two strands composing the polymer come in contact in such a way as to form a 
bound double structure, which remains swollen and does not assume compact 
configurations. 
In order to induce a transition from unzipped to zipped state, the minimal 
inhomogeneity required implies a distinction between the two strands: if the 
polymer is made of two blocks composed of monomers of different species
(diblock copolymer) and there is a dominant attractive interaction acting 
between these different monomers, one would expect such a transition to be 
possible. Of course, the zipping occurring in biomolecules results in general 
from higher degrees of inhomogeneity than those of a simple diblock copolymer.

From a physical point of view one can think to diblock copolymers with 
oppositely charged monomers in the two blocks; in the model discussed here 
the interactions are of short range, and this would correspond to the case 
of screened Coulomb forces. Another possibility is that the attractive 
interactions between monomers of the two blocks are established through a 
preferential formation of hydrogen bonds. 
The attractive interactions between the two blocks, besides zipping, tend also
to produce collapse into a globular compact state, unless some contrasting 
effect limits the capability of a given monomer to attract monomers of the 
other block.

In a recent paper \cite{OSS00}, a model of diblock copolymer with some of 
the features discussed above has been studied in both two dimensions (2D) 
and 3D. In that model the two blocks were represented by two halves of a
self avoiding walk (SAW) on hypercubic lattice, and attractive interactions 
were acting between nearest neighbor sites (monomers) visited by the two 
blocks.
So, apart from the steric constraints, there was no interaction mechanism
possibly opposing the tendency of a given monomer to be surrounded 
by as many as possible monomers of the other block. 
The transition of the diblock copolymer from a high T swollen to 
a low T compact phase, had analogies with both polymer adsorption 
on a wall, and $\Theta$-collapse, but turned out to belong to a universality 
class different from both \cite{OSS00}.
An intriguing question remained open concerning the very nature of this 
transition: indeed, the possibility that a zipped, swollen phase could exist 
for temperatures just below the transition, could not be excluded. 
If this were the case, the adsorption-like collapse found in Ref. \cite{OSS00}
would correspond to a zipping and a further transition to the compact globular 
phase should be expected to take place at a lower T. 

In the present article we extend the model of Ref. \cite{OSS00} in 2D to
include an interaction among alternating triplets of different monomers.
Depending on its sign, this additional interaction can enhance the tendency of
the system either to form compact structures, or to take zipped conformations.
We draw an accurate phase diagram for the system, in which a zipping 
transition line is well-identified and characterized. Our analysis seems to
indicate that the adsorption-like collapse of Ref. \cite{OSS00} belongs to
the zipping universality class as well.
Specifically, we find that, depending on the triplet interaction energy one 
has either a continuous swollen-zipped transition followed by a first order 
collapse into compact conformations, or a direct first order swollen-compact 
transition.
Although we mainly focus on the zipping, it turns out that also the first 
order collapse has interesting features, since it shows remarkable analogies 
with that found in homopolymers with orientation dependent interactions, which 
attracted some attention recently \cite{oriented}.

We will argue that the exact exponents for the zipping transition in 2D can 
be found through an identification between the stochastic geometry of the 
blocks and that of a percolation cluster backbone \cite{percobook}. 
A preliminary form of this argument was presented in Ref. \cite{OSS00}. 
The relevant dimensions of the percolation cluster can be identified thanks 
to some recent results for path-crossing probabilities \cite{ADA99}. Our 
numerical estimates for the zipping exponents are in very good agreement 
with the conjectured values.
The mapping onto percolative stochastic geometry we discuss here
is, to our knowledge, the first example of exact results derivation 
for a genuinely inhomogeneous polymer problem in 2D. 
The connection with percolation geometry shows also that the physics of zipping 
is closely connected with that of the $\Theta$-point transition, for which 
a representation in terms of percolation geometry has been established already 
long ago \cite{percolation}. 

Besides the prototypical importance that zipping acquires here also on the 
basis of our exact results, one should realize that many of the conformational 
transitions occurring in biomolecules show aspects which, to some extent, are
reminiscent of zipping. This is certainly the case for DNA, in which, upon 
lowering temperature below the denaturation one, the conjugate bases form 
pairs, so that the molecule arranges itself in a double stranded helical 
structure \cite{denaturation}. 
Double stranded, zipped structures appear also in the folding of 
$\beta$-hairpin peptides \cite{betahairpin}.
In the philosophy mentioned at the beginning of this section, one can hope 
that studies like the one presented here can teach something about how to 
model properly these more complicated systems.

This paper is organized as follows: In Section \ref{sec:pd} we present the 
model and the main features of the phase diagram. In Sections \ref{sec:exact} 
and \ref{sec:mc} we discuss the numerical results obtained by exact 
enumerations and Monte Carlo simulations, respectively.
In Section \ref{sec:perco}, using recent results for crossing probabilities
of percolation paths in 2D, we conjecture the exact values of the exponents
of the zipping transition.
Section \ref{sec:conclusions} concludes the paper, with a summary of the
results and a general discussion. 

\section{The Model and Phase diagram}
\label{sec:pd}

We model the 2D diblock copolymer by an interacting SAW on the square lattice.
In the configuration $w$ the SAW has $|w| = N$ vertices (monomers), with $N$
even, and consists of $N/2$ consecutive monomers of type A ($w_A$) followed 
by $N/2$ monomers of type B ($w_B$). A pair of vertices $(A,B)$ form a contact 
if they are a unit lattice distance apart.
The interaction between the two blocks $w_A$ and $w_B$ is taken into account 
by assigning an energy $\varepsilon$ ($\varepsilon <0$) to each $A-B$ contact.
In addition we introduce a second energy parameter $\delta$, associated with 
contacts formed by a sequence A-B-A or B-A-B of neighboring monomers on a line.
We refer to these sequences as to {\it triple} contacts.
The Hamiltonian of the system in configuration $w$ is given by:
\begin{equation}
H(w) = N_{AB}(w) \varepsilon +  N_{3}(w)\;\delta
\end{equation}
where $N_{AB}(w)$ and $N_{3}(w)$ are the number of $A-B$ and of triple 
contacts, respectively. 

For $\delta =0$ we recover the model introduced in Ref. \cite{OSS00};
in the present work we consider both signs of $\delta$: a positive value of 
$\delta$ must prevent the polymer from collapse into compact conformations 
and favor an intermediate zipped phase, while for a negative $\delta$ the 
tendency to collapse is enhanced.

\begin{figure}[b]
\centerline{
\psfig{file=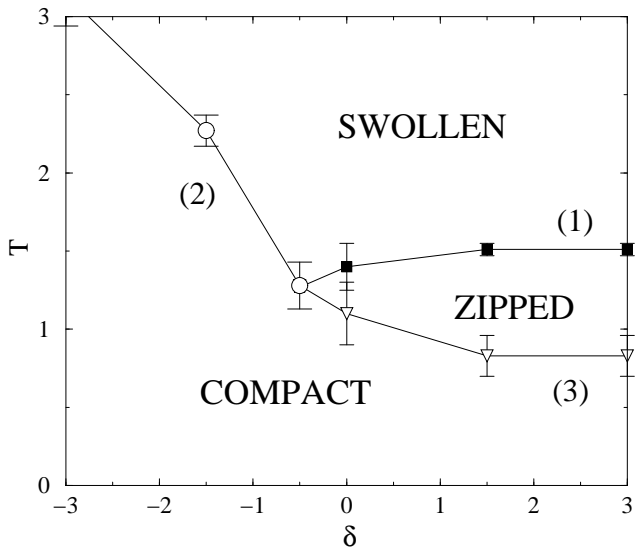,height=7.5cm}}
\vskip 0.2truecm
\caption{Phase diagram in the $\delta$-$T$ plane. Line (1) is the continuous 
zipping transition line separating the high temperature swollen phase from 
the zipped phase. The lines (2) and (3) are of first order type.
}
\label{FIG01}
\end{figure}

Letting $c_{N}(N_{AB},N_{3})$ be the number of copolymer configurations with 
$N$ edges, $N_{AB}$ contacts of type $A-B$ and $N_{3}$ triple contacts, we 
define the finite-$N$ free energy per monomer
\begin{equation}
F_{N}(\beta,\delta) = N^{-1} \log {Z_{N}(\beta,\delta)},
\end{equation}
where
\begin{equation}
Z_{N}(\beta,\delta) = 
\sum_{N_{AB},N_{3}} c_{N}(N_{AB},N_{3}) \,\,\, 
e^{-\beta (N_{AB} \varepsilon + N_{3} \delta)}
\end{equation}
is the partition function and $\beta = 1/T$ \cite{convergence}. Throughout 
the rest of the paper we set $\varepsilon = -1$. By varying $T$ and
$\delta$ we have explored the phase diagram of the model (see 
Fig. \ref{FIG01}) on the basis of exact enumeration and Monte Carlo 
simulation results.

As expected, we find a zipped phase in the positive $\delta$ region, while 
for negative $\delta$ there is a direct transition from the swollen to the 
compact phase. The numerical results show that the line (1) separating the 
swollen and zipped phases is continuous, while (2) and (3) are first order.
Unfortunately, the numerical methods at our disposal are not sufficiently
accurate to determine the precise location of the intersection point between
the lines, nor the character of the phase transition in this point, which
could be of special type. The location of this intersection point seems to 
fall at slightly negative values of $\delta$.

\begin{figure}  
\centerline{
\psfig{file=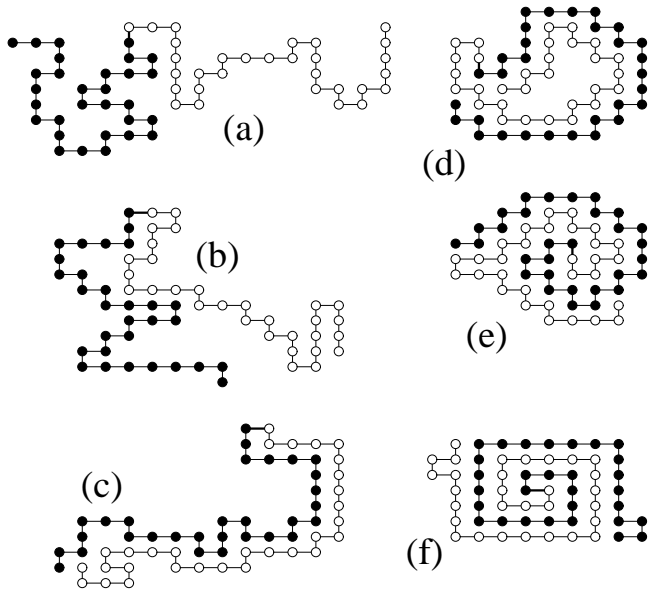,width=8.5cm}}
\vskip 0.2truecm
\caption{Typical Monte Carlo equilibrium configurations for an $N=60$ diblock 
copolymer: a-b) swollen chains; c-d) zipped chains; e-f) compact chains with 
$\delta=1.5$ (e) and $\delta=-1.5$ (f).} 
\label{FIG02}
\end{figure}

It is rather instructive to show some typical equilibrium conformations of
the copolymer for various values of $T$ and $\delta$, in 
the three different phases (see Fig. \ref{FIG02}). The configurations are
snapshots obtained by Monte Carlo simulations: (a) and (b) are 
conformations in the swollen high $T$ phase, with (b) close to the 
zipping transition. 
(c) and (d) are instead zipped configurations, with (d) sampled in the vicinity
of the transition to the compact state; notice that the pairing of the two 
strands in (d) follows an opposite orientation with respect to (c).
Finally (e) and (f) are both compact, but of different nature: the latter 
occurs at $\delta < 0$ where triple contacts are energetically favored. 
Therefore the polymer assumes a spiral-like shape
with straight segments turning around the center in order to maximize the
number of triple contacts. In the case (e) $\delta$ is positive:
the configuration is still of spiral type, but in this case the arms of the
spiral are oriented preferentially at $45^o$ with respect to the axes of
the square lattice, in order to avoid the formation of triple contacts.

One can easily understand now why the lines (1) and (3) for large $\delta$ 
run practically horizontal in the phase diagram of Fig.~\ref{FIG01}.
In the whole zipped phase triple contacts, which cost an energy $\delta$, 
seldom occur (see Fig.~\ref{FIG02}(c)), therefore the zipping temperature 
should depend rather weakly on $\delta$. In addition the polymer can form 
compact conformations such as that shown in Fig. \ref{FIG02}(e), which also 
avoids this type of contacts. Hence, also the zipped-compact transition 
temperature should not depend on $\delta$, when $\delta$ is positive and large 
enough. Both lines (1) and (3) in Fig.~\ref{FIG01} should be asymptotically 
horizontal for large $\delta$.

\section{Exact enumerations}
\label{sec:exact}

Exact enumerations of interacting SAW's are standard techniques for the study 
of the homopolymer $\Theta$-collapse transition \cite{series}.
In the present calculation we generated all possible configurations for 
copolymers up to $N=30$ monomers, a length which is already sufficient to 
characterize rather well the critical behavior of the zipping transition.

The occurrence of phase transitions in interacting polymer systems can be 
detected by studying the large $N$ behavior of the canonical average 
squared radius of gyration
\begin{equation}
R_g^2 = \frac{\sum_{w} \exp [-\beta H(w)] \,\, R^2(w)} 
{\sum_{w} \exp [-\beta H(w)]},
\label{radius}
\end{equation}
where the sums extend to all $N$-steps configurations, $w$, of the copolymer, 
with radius $R(w)$ relative to the center of mass. Indeed, in the proximity 
of a conformational transition temperature $T_c$ we expect that
\begin{equation}
R_g (N,T)  \sim N^{\nu_c} {\cal R}\left [(T-T_{c})N^{\phi}\right ]
\label{scalr}
\end{equation}
where $\nu_c$ and $\phi$ are the exponents characterizing the transition and 
${\cal R}$ is a scaling function that is assumed to approach a positive
constant if its argument approaches zero. 
 
Another important quantity is the specific heat $C(N,T) = 
\frac{1}{N}\partial \langle H\rangle /\partial{T}$, which for $N$ large
and $T$ close to $T_{c}$ is expected to obey the scaling:
\begin{equation}
C (N,T) \sim N^{2\phi -1}{\cal C}\left [(T-T_{c})N^{\phi}\right ]
\label{scalc}
\end{equation}
where ${\cal C}$ is again a suitable scaling function.

\begin{figure}[b]
\centerline{
\psfig{file=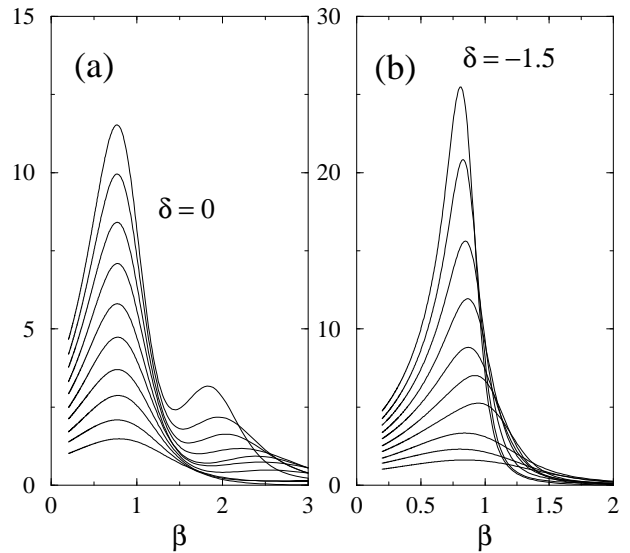,height=7.5cm}}
\vskip 0.2truecm
\caption{Derivatives of the squared radius of gyration with respect to the 
inverse temperature for $\delta = 0$ (a) and $\delta = -1.5$ (b) and for 
$N=12, 14, \ldots 30$. The double peak structure for $\delta = 0$ suggests 
the sequence of two transitions, swollen-zipped and zipped-collapsed.}
\label{FIG03}
\end{figure}

\begin{figure}[b]
\centerline{
\psfig{file=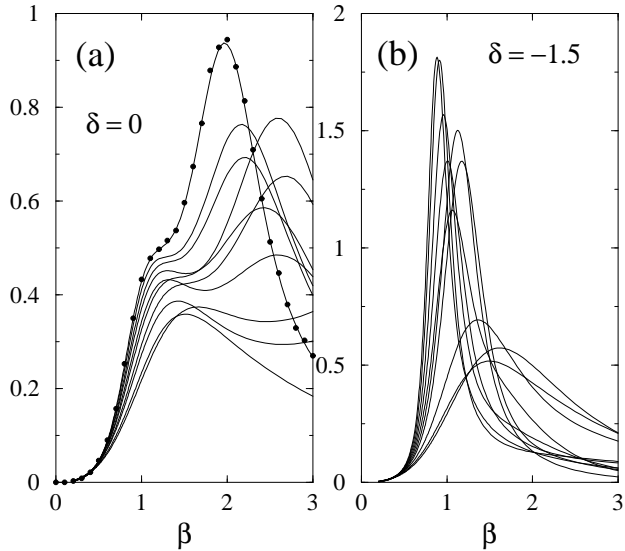,height=7.5cm}}
\vskip 0.2truecm
\caption{Solid lines: Specific heat from exact enumerations for $N=12$, $14$,
\ldots $30$ for $\delta = 0$ (a) and $\delta = -1.5$ (b). Circles: Monte Carlo
results for $N=30$.}
\label{FIG04}
\end{figure}

Figures \ref{FIG03} and \ref{FIG04} show $d{R_g^2}/d\beta$ and $C$ as 
functions of $\beta$ for $\delta =0$ (a) and $\delta = -1.5$ (b).
As the radius of gyration drops at a transition, its 
derivative $d{R_g^2}/d\beta$, shows a peak in correspondence to the 
transition point.

For $\delta = -1.5$ both quantities have a single isolated peak, indicating 
that there is a single transition from a swollen to a compact phase.
For $\delta =0$, instead, the derivative of the radius of gyration has two
distinct peaks (Fig.~\ref{FIG03}(a)), while the picture emerging from the
specific heat (Fig.~\ref{FIG03}(b)) is somewhat more confusing, since the 
$N$-dependence of the peak positions and heights is rather irregular and
their extrapolation to $N \to \infty$ becomes impossible.
Therefore we focus on the peaks of $dR_g^2/d\beta$. Extrapolating their 
positions at $\delta=0$ we finds the following two estimates of critical 
temperatures $T_c = 1.40(15)$ and $T_{2c} = 1.1(20)$, which overlap somewhat 
within error bars. For this reason it is difficult to discern between two 
separate, but close, transitions, and a single one. 
However, as $\delta$ is increased, the two sets of peaks get clearly separated 
and extrapolations yield two distinct transition temperatures. We focus here 
on the characterization of the high T transition from the swollen to the zipped 
phase.
The scaling form of Eq. (\ref{scalr}) implies that $T_c(N)$, the temperature
where $d{R_g^2}/d\beta$ has a maximum, scales for large $N$ as 
$T_c(N) - T_c \sim x_0 N^{-\phi}$, with $x_0$ a suitable constant.
We calculated both the radius of gyration and specific heat at $T_c(N)$;
from (\ref{scalr}) and (\ref{scalc}) one has:
\begin{equation}
R_g (N,T = T_c (N)) \sim N^{\nu_c} \,\, {\cal R}(x_0)
\end{equation}
and
\begin{equation}
C (N,T = T_c (N)) \sim N^{2 \phi-1} \,\, {\cal C}(x_0)  
\end{equation}
For the calculation of the critical exponents we formed first the finite 
$N$ approximants, for instance 
\begin{equation}
\nu_c (N) \equiv \frac{\ln(R_g (N+2)/R_g(N))}{\ln((N+2)/N)}
\end{equation}
(here $R_g(N)$ is a shorthand notation for $R_g(N,T_c(N))$) and then 
extrapolated $\nu_c(N)$ to $N \to \infty$. The same procedure was followed 
for $\phi$.

The extrapolated values are reported in Table \ref{TABLE01}, together with the 
exponent $\nu'_c$, which is that associated to the radius of the half-chain,
or single block, which at the critical temperature should scale as:
\begin{equation}
R'_g (N, T=T_c) \sim \left( \frac{N}{2} \right) ^{\nu'_c}
\end{equation}

\vbox{
\begin{table}[tb]
\caption{Extrapolated values of $\nu_c$ and $\phi$ from the exact enumeration 
data relative to diblock copolymers up to length $N=30$. The exponent $\nu'_c$ 
is obtained from the scaling behavior of the radius of one of the blocks.} 
\label{TABLE01}
\vskip 0.2truecm
\begin{tabular}{cccc}
$\delta$ & $\nu_c$ & $\phi$ & $\nu'_c$ 
\\ \hline
0.0 & 0.72(1) & 0.60(5) & 0.74(1)\\
0.5 & 0.73(1) & 0.58(3) & 0.750(5)\\
1.0 & 0.73(1) & 0.57(3) & 0.750(5)\\ 
3.0 & 0.74(1) & 0.56(3) & 0.750(5)
\end{tabular} 
\end{table}
}

The values of the exponents $\phi$ and $\nu_c$ vary slightly along the line (1)
when $\delta$ is increased, while $\nu'_c$ is rather stable. We believe that
the variation of $\nu_c$ and $\phi$ is a spurious effect due to the vicinity 
of an additional transition in the neighborhood of $\delta=0$. 
It is much more plausible that the exponents are constant along the line (1); 
the most reliable estimates for $\nu_c$ and $\phi$ should be those for large 
$\delta$, where the lines (1) and (3) are clearly separated.
The values of $\nu_c$ and $\nu'_c$ are consistent (the former only at large 
$\delta$) with the scaling behavior of a SAW, namely, $R_g \sim N^{3/4}$.
The value of $\phi$ is instead consistent with $\phi = 9/16 = 0.5625$, which 
was conjectured in Ref. \cite{OSS00} for the transition at $\delta=0$ and 
will be derived in detail in Section \ref{sec:perco}.

As for the transition lines (2) and (3), the exact enumeration analysis is
not at all conclusive since the scaling behavior of the peaks with the chain 
length $N$ is rather irregular and precise extrapolations turn out to be 
impossible.
This issue will be clarified with the use of Monte Carlo simulations, which 
allow to achieve much larger copolymer lengths.
      
\section{Monte Carlo simulations}
\label{sec:mc}

In order to sharpen and extend the results of exact enumerations, in
particular concerning the properties of the zipped and collapsed phases, we 
performed Monte Carlo simulations for various $N$ and $\delta$ and for a 
wide range of temperatures. 
Since the simulations considered involve sampling at points which include
low values of $T$, a standard Markov chain Monte Carlo approach is unlikely 
to be successful, being difficult to construct a Markov chain sufficiently 
"mobile" at low $T$ where the interaction energies become relevant.
Instead we use a multiple Markov chain technique by which one samples 
simultaneously at various values of $T$, including $T=\infty$ where converge 
is rapid. Most recently this method has been used successfully to investigate 
collapse transitions in homopolymers \cite{TJOW96}, hetheropolymers 
\cite{monari}, and adsorption in $\Theta$-solvent \cite{VW98}.

First one defines a Metropolis based Markov chain for a temperature $T$.
This procedure makes use of a hybrid algorithm based on pivot \cite{MS88} 
as well as on local moves \cite{VS61}.       
Pivot moves are of global type and operate well in the swollen regime, whereas 
local moves turn out to be essential in speeding up Monte Carlo convergence at 
low temperatures \cite{TJOW96}. In our calculations, each Monte Carlo step
consists of O(1) pivot moves and O(N) local moves. In this model, however, we 
have to deal also with a zipped phase where the most probable configurations 
are characterized by having the two blocks A and B paired together, but still 
not compact (see Fig. \ref{FIG01} (b,c,d)). 
To increase the mobility of the Markov chain in this region we added a set of 
bilocal moves, such as end-end reptation and kink-end (and end-kink) moves 
\cite{CCFPP}. These moves are particularly effective for dense chains, and 
even more effective for the zipped chains, where typically one side of each 
half-chain is free and can hold a new kink. The resulting algorithm is a 
little heavier, but enables the reciprocal sliding of the half-chains and a 
more efficient exploration of the configuration space.                              
One may then run in parallel a number $m$ (typically $20-40$) of these Markov 
chains at different temperatures. The sampling at low $T$ is then considerably 
enriched by swapping configurations between Markov chains contiguous in $T$. 
The whole process is itself a (composite) Markov chain, obeys detailed balance 
and is ergodic \cite{TJOW96}.  

Monte Carlo simulations were performed for three distinct values of $\delta$:
$\delta=1.5$, $\delta=-1.5$ and $\delta=0$. As a test of the performance of the
multiple Markov chain algorithm we compared the Monte Carlo results with those 
obtained from the exact enumeration for chains up to $N=30$ monomers. In all
cases analized the agreement turned out to be extremely good (see, for 
instance, Fig. \ref{FIG04}(a)).

\subsection{$\delta = 1.5$}

\begin{figure}   
\centerline{
\psfig{file=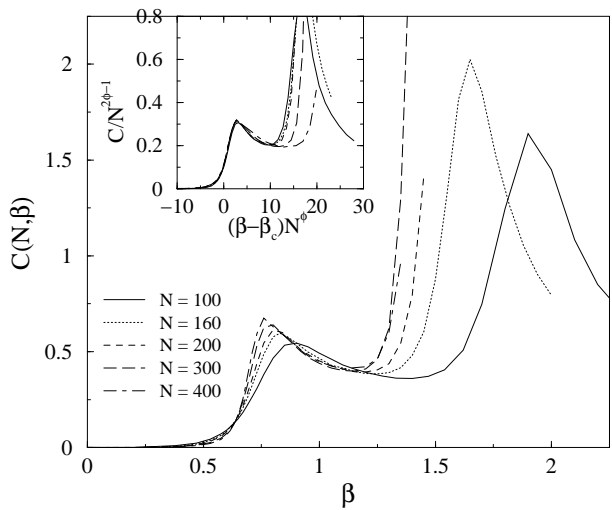,width=8.cm}}
\vskip 0.2truecm
\caption{$\delta = 1.5$: Plot of the specific heat vs. $\beta$, for chains
of various lengths. For $N=300$ and $N=400$ the low temperature peak was not 
reached because of the considerable autocorrelation time caused by low mobility
of long collapsed chains. Inset: collapsed specific heat $C/N^{2\phi-1}$ vs. 
$(\beta-\beta_c)N^\phi$, with $\phi= 9/16 = 0.5625$ and $\beta_c = 0.66$.} 
\label{FIG05}
\end{figure}

In the case $\delta = 1.5$ we considered diblocks of lengths up to $N=400$ 
and sampled at a set of $m\simeq40$ temperatures typically in the range 
$T \in [0.5,\infty]$.
In Fig.~\ref{FIG05} we plot the specific heat as a function of $\beta$ for 
different $N$ values. Clearly each curve displays a double peak structure 
indicating two subsequent transitions. We can rule out the possibility that 
such double peaked structure is a finite size effect 
by noting that the peaks sharpen and grow with $N$. 
Let us focus first on the set of peaks at higher temperatures, i.e. on the 
transition from a swollen to a zipped phase. 
The corresponding $T_{c}$ and $\phi$ could be deduced from the $N$ dependence 
of the height, $h(N)$, and position, $T_{c}(N)$, of the peak maxima. 
Indeed, from the scaling behavior (\ref{scalc}) we expect, as $N$ increases,
\begin{equation}
h(N) \sim N^{2\phi-1} \quad \hbox{and} \quad T_{c}(N) - T_{c} \sim N^{-\phi}
\end{equation}                                                    
Since a linear least squares fit of $\log h$ vs. $\log N$ gives a very large 
$\chi^{2}$ 
statistical error, we fit the data with a function $A\,N^{2\phi-1}(1+B/N)$ 
where a scaling correction $1/N$ is included. The least squares fit in this 
case gives $\phi=0.57\pm 0.02$, in agreement with the value $\phi=9/16=0.5625$
conjectured in Ref.~\cite{OSS00} and also with the estimates obtained by exact 
enumeration. This procedure yields results consistent with a direct 
extrapolation of effective finite $N$ exponents.
We have also tried to fit the data by using the more general scaling correction
$1/N^\Delta$, but the best fit is obtained for $\Delta =1$. 
The estimated value of $\phi$ allowed us to extrapolate $T_{c}$ by plotting 
$T_c(N)$ vs. $1/N^\phi$. This gives $T_c=1.51(4)$ ($\beta_c=0.66(2)$).
                                                
The inset of Fig. \ref{FIG05} shows a plot of $C/N^{2 \phi -1}$ vs. 
$(\beta - \beta_c) N^\phi$, where we have used $\phi = 9/16$ and the
estimate $\beta_c = 0.66$.
As expected, the  high temperature peaks collapse onto a single curve quite 
nicely. On the contrary, for the set of peaks at lower temperatures, the same
rescaling procedure turns out to be inappropriate. In particular, by using 
the rescaled variables adequate for the former set of peaks, the positions 
of the latter set tend to move away from zero while their heights still 
increase with $N$: this is only consistent with a scenario in which a new 
transition, at a lower $T=T_{{2c}}$, exists. 
This transition should be also characterized by a crossover exponent greater 
than $\phi=9/16$. We have tried to verify this by looking for two new values 
$\beta_{2c}$ and $\phi_{2}$ that allow a reasonable fit of the scaling 
behavior of the second set of peaks. In this way
we obtained the rough estimates $\beta_{2c} \approx 1.0$ and $\phi_{2}\approx 
0.7 >9/16 $. Unfortunately the sampling at low temperatures is not 
good enough to make such estimates sufficiently sharp. Moreover, in the zipped 
phase the effective size of an $N$ monomers system drops to $N/2$, making the 
finite size corrections to scaling more pronounced.

\begin{figure}   
\centerline{
\psfig{file=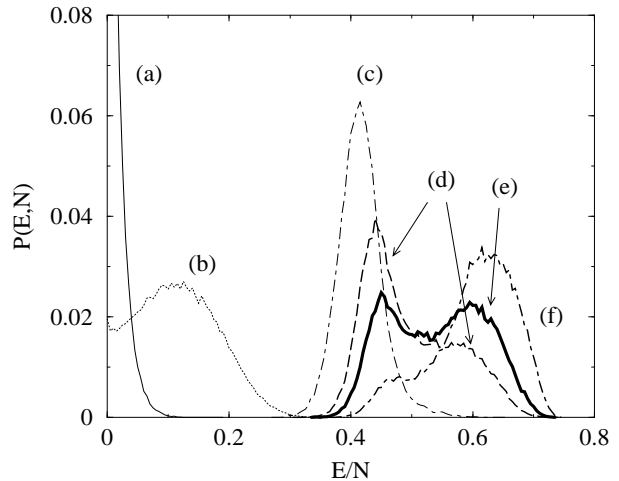,width=8.cm}}
\vskip 0.2truecm
\caption{Plot of $P(E,N)$ with $N=200$ and for various temperatures: 
(a) $\beta = 0.3$, (b) $\beta = 0.75$, (c) $\beta = 1.4$, (d) $\beta = 1.58$,
(e) $\beta = 1.66$ and (f) $\beta = 1.8$.
}
\label{FIG06}
\end{figure}

The different nature of the two transitions can be better detected from the 
behavior of $P(E,N)$, the probability distribution of the energy $E$, for a 
chain of length $N$. 
Figure \ref{FIG06} shows a plot of $P(E,N)$ as a function of $E/N$ for $N=200$. 
At sufficiently high temperatures this quantity has a maximum in $E=0$ and 
decreases rapidly with $E/N$ (Fig. \ref{FIG06}(a)). 
As the temperature is lowered the maximum shifts continuously to larger 
values of $E/N$ (b,c). For lower temperatures $P$ develops a double peak
structure (d,e,f). In the case (e) the peaks have equal height, while at 
temperatures below or above it one of the two peaks dominates over the other.
This behavior, which persists and becomes more pronounced upon increasing 
$N$, is an indication of phase coexistence; hence the transition at lower $T$ 
is of first order type. In terms of specific heat this would mean $C(T) \sim 
N$ as $T = T_{2c}$, i.e. $\phi_{2} = 1$. By extrapolating $\beta_{2c}(N)$ vs. 
$1/N$ we find $\beta_{2c} = 1.2\pm 0.2$. 

Another way to characterize the different phases of the model consists in 
looking at the scaling behavior of metric quantities such as $R_g^2$ defined 
in (\ref{radius}) and the mean squared end-to-end distance ${R^2_e}(N) = 
\langle (r_{N}-r_{0})^{2}\rangle$, where $r_0$ and $r_N$ are the two 
end-monomers of the copolymer.
For large $N$ we expect
\begin{eqnarray}
  {R^2_e}(N) &\sim& \rho_{e}N^{2\nu} \label{Re} \\
  {R^2_g}(N) &\sim& \rho_{g}N^{2\nu} \label{Rg}
\end{eqnarray}
and an interesting quantity to be computed is the ratio $\rho_{e}/\rho_{g}$, 
which is expected to be universal \cite{domb14}.
For non interacting SAW's on a square lattice, exact enumerations and Monte
Carlo simulations give $\rho_{e}/\rho_{g} \sim 7.13$ (see references in
\cite{domb14}). Figure \ref{FIG06} shows the ratio ${R^2_e}/{R^2_g}$ as a 
function of $\beta$ for several $N$. Note that in the range of $0<\beta < 
0.66$ the curves tend to assume a constant value $\rho_e/\rho_g=7.15(5)$ in 
agreement with the value expected for non interacting SAW's. In the proximity 
of the zipping transition the curves start to bend downwards and at 
$\beta_{c} \approx 0.66$ they cross each other almost in a 
unique point (see inset). At the crossing point our estimate of the 
universal amplitude is $\rho_e/\rho_g = 6.35\pm0.20$ which is definitively 
different from the amplitudes ratio of the SAW universality class. 
The zipping transition that cannot be 
distinguished from the swollen phase in terms of the $\nu$ exponent, is 
however characterized by a different value of the universal ratio 
$\rho_e/\rho_g$ \cite{notagg}.

\begin{figure}   
\centerline{
\psfig{file=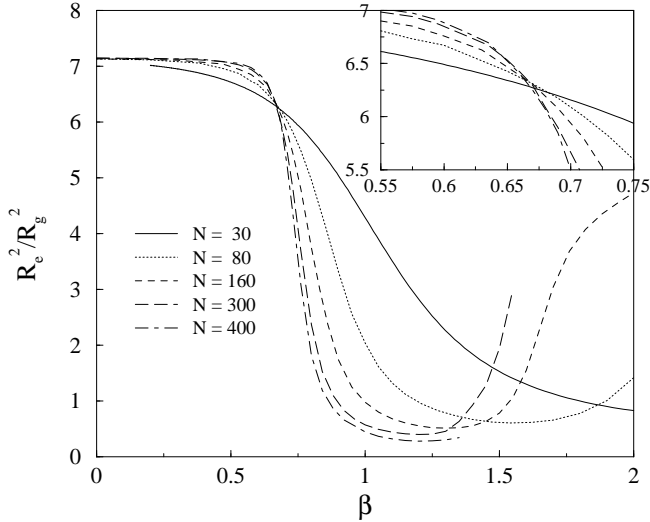,width=8.5cm}}
\vskip 0.2truecm
\caption{$R_e^2/R_g^2$ vs. $\beta$ for $\delta = 1.5$ and for various chain 
lengths. The $N=30$ curve is calculated from exact enumerations, while the 
others are obtained from Monte Carlo simulations. Inset: blow up of the
crossing region.}
\label{FIG07}
\end{figure}
 
If the temperature is further lowered the $R^2_e/R^2_g$ curves reach a 
minimum value that decreases as $N$ increases. This is an indication that 
the end-to-end distance in the zipped phase does not scale anymore like 
the radius of gyration, as assumed in Eqs. (\ref{Re},\ref{Rg}).
For sufficiently low temperatures, $\rho_{e}/\rho_{g}$ starts to grow back 
indicating that the compact phase is characterized by end-to-end distance 
and mean radius of gyration that scale in the same way with $N$. As the typical
low $T$ configurations are of spiral type (see Fig. \ref{FIG02}(e,f)) with 
end-points at opposite sides of the spiral, it is natural to expect that 
$R_e \sim R_g \sim N^{1/2}$.

\subsection{$\delta = -1.5$}

\begin{figure}   
\centerline{
\psfig{file=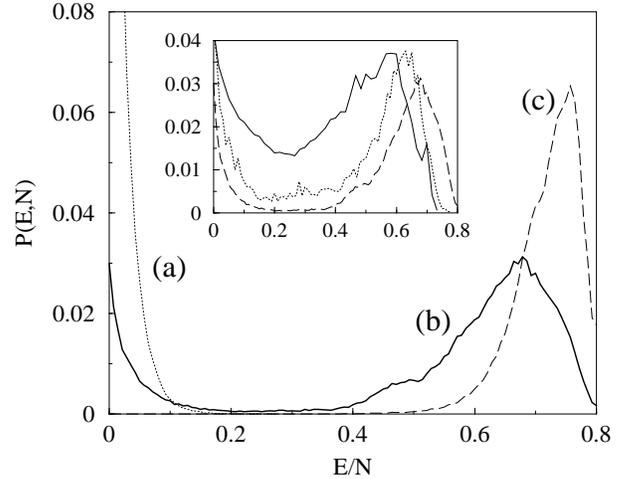,width=8.cm}}
\vskip 0.2truecm
\caption{$P(E,N)$ vs. $E/N$ for $\delta = -1.5$ and $N=200$ for three different 
temperatures (a) $\beta = 0$, (b) $\beta = 0.545$ and (c) $\beta = 0.6$.
Inset: Plots of $P(E,N)$ vs. $E/N$ near phase coexistence for $N=60$,
$\beta = 0.66$ (solid line), $N=100$, $\beta = 0.56$ (dotted line) and $N=140$,
$\beta = 0.545$ (dashed line).
}
\label{FIG08}
\end{figure}

For $\delta<0$ triple contacts are favored and we expect (as the exact 
enumerations already indicate) a single transition from swollen directly 
to compact phase.
To investigate the nature of such transition we have performed runs with 
$\delta = -1.5$, for several values of $N$, sampling at $m\simeq 30$ 
different temperatures in the interval $T \in [1.3,\infty]$. As in the case 
$\delta =1.5$ we have examined the probability of finding the copolymer in 
a configuration with energy $E$, as a function of the temperature. 
A plot of $P$ for $N=200$ and three different temperatures is shown in 
Fig.~\ref{FIG08}. 
Close to the transition temperature $P$ has two maxima 
(see Fig.~\ref{FIG08}(b)) one at $E=0$ and the other at $E/N \approx 0.7$.
This is a clear indication of a first order transition.          
The evidence of such a behavior is stronger than in the case $\delta =1.5$ 
since here the coexistence is between two phases (the swollen and the compact)
with a rather large difference in energy, therefore the double peak structure 
of $P$ can be noticed already for small $N$.

From the analysis of the specific heat peaks we find that they become sharper 
as $N$ increases and their height appears to grow with a power of N slightly
exceeding one (the physical upper limit). At the same time, in a plot of energy 
vs. temperature we see curves that seem to approach step functions. These data 
support the idea that the corresponding transition should be of first order. 

\subsection{$\delta =0$}

The more delicate region to be explored in the phase diagram is the
neighborhood of $\delta = 0$, where three transition lines meet each 
others. We have chosen in particular the case $\delta = 0$, 
since it was considered in Ref. \cite{OSS00}.

\begin{figure}   
\centerline{
\psfig{file=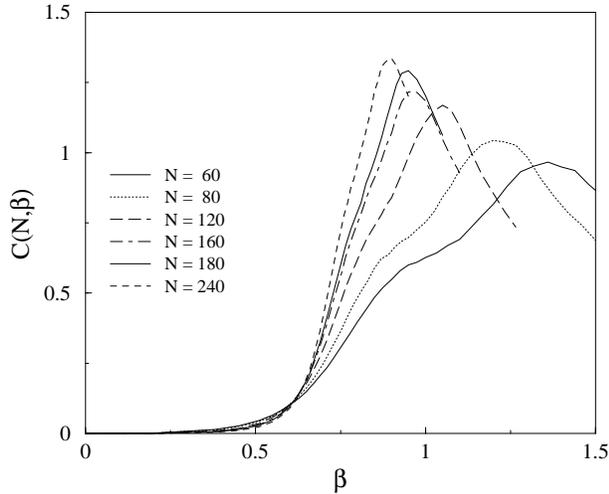,width=8.cm}}
\vskip 0.2truecm
\caption{$\delta = 0$: Plot of the specific heat vs $\beta$, for various chain
lengths.}
\label{FIG09}
\end{figure}

In Fig.~\ref{FIG09} we plot the specific heat as a function of $\beta$ for 
several $N$ values. For the smallest chains ($N =60, 80$), one observes a 
peak in the specific heat with a shoulder at smaller $\beta$.
When the copolymer length is increased the shoulder becomes hardly noticeable.
From the specific heat plot one cannot rule out the possibility that the
shoulder eventually vanishes leaving out a single transition from a swollen 
to a compact phase. The other possibility is that there are two separate 
transitions, but very close in temperature.

\begin{figure}   
\centerline{
\psfig{file=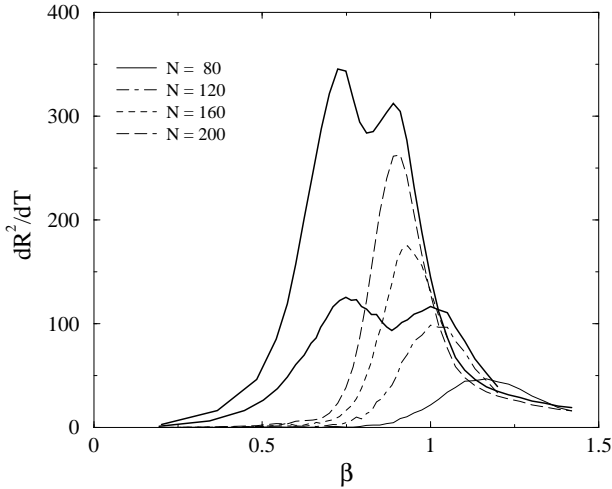,width=8.cm}}
\vskip 0.2truecm
\caption{$\delta = 0$. Thin lines: plot of the derivative of the half-chain
squared gyration radius, as function of $\beta$. Thick
lines: two examples of derivative of the total-chain squared gyration radius
(N=120, 200).}
\label{FIG10}
\end{figure}

The presence of two distinct transitions is suggested by a
plot of the temperature derivative of the total radius of gyration, shown
for $N=100$ and $N=200$ as thick lines in Fig. \ref{FIG10}.
In this case one clearly detects two peaks, which although coming closer
to each other as $N$ increases are still noticeable and sharp for $N$ rather 
large. The thin lines in Fig. \ref{FIG10} are the temperature derivatives of 
the radius of gyration of a single block, which show only one peak in 
correspondence to the low temperature peak of the derivative of the total 
radius of gyration.
This behavior is consistent with the following picture: coming from the swollen
phase (small $\beta$) one has first a zipping transition characterized by a
drop of the total radius of gyration, while the radius of gyration of a single
block still behaves as a SAW and is not sensible to the zipping transition.
However, at lower temperatures, in correspondence to the transition from zipped
to compact phase both quantities drop and their derivatives show a peak.

\begin{figure}   
\centerline{\psfig{file=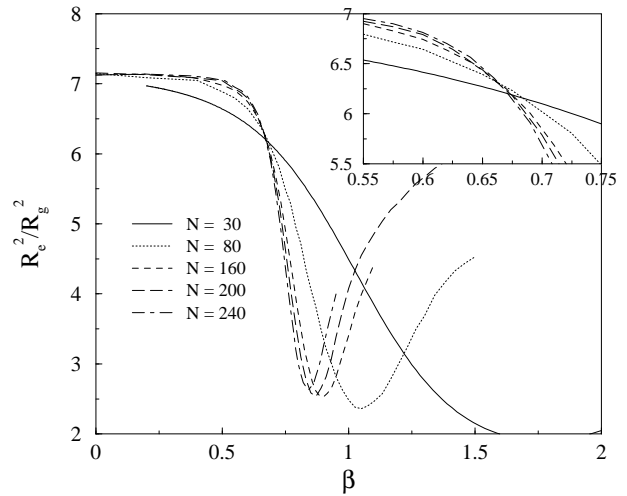,width=8.cm}}
\vskip 0.2truecm
\caption{$\delta = 0$: Square end-to-end distance over square gyration 
radius as a function of $\beta$.}
\label{FIG11}
\end{figure}

Another quantity which we investigated is the universal amplitude ratio between
the end-to-end distance an radius of gyration squared, which is plotted in
Fig. \ref{FIG11}.
Here, as for the $\delta = 1.5$ case, this universal quantity takes the SAW 
value $\sim 7.13$ at high T and drops in correspondence of the transition. 
The fact that we find intersections with $R^2_e/R^2_g \approx 0.635$ (the 
same value as for $\delta = 1.5$) strongly suggests the presence
of a zipping transition with the same universal properties as that 
at $\delta = 1.5$. Unlike from Fig. \ref{FIG07} here $R_e^2/R_g^2$ drops and
increases again in a narrow range of $\beta$ values indicating the the 
zipped phase is restricted to a small temperature interval.

In summary, although the numerical evidence is not fully conclusive, our data
seem to favor the existence of two separate transitions for $\delta=0$.
As in the case $\delta = 1.5$, it is natural to expect that the low T one 
(zipped-collapsed) is of first order type.

\section{Percolation paths and exact exponents of the zipping transition}
\label{sec:perco}

In this section we present a conjecture on the relation between the
statistics of some percolation paths at threshold and the diblock copolymer
zipping transition. This conjecture leads to predict exact values for the 
exponents. A preliminary, less precise version of the arguments below was 
given in Ref. \cite{OSS00}.

It is well-known that, in 2D, the statistics of a ring polymer at the
$\Theta$-transition is identical to that of the external perimeter, or hull,
of a percolation cluster.
Through this identification the exact exponents of the $\Theta$-transition, 
$\nu_\Theta = 4/7$ and $\phi_\Theta = 3/7$, were derived \cite{percolation}.
Here, we show how similar arguments can be invoked for the zipping transition.
The differences are mainly associated to the fact that the relevant percolative
set appropriate for the zipping is not the hull, as for the homopolymer 
$\Theta$-point.

Like in the $\Theta$-point case, here it is convenient to consider site
percolation on a triangular lattice. For this problem the relevant percolation
contours, like the hull of a cluster, are in fact strictly self-avoiding paths
on the dual, hexagonal lattice. Thus, also the equivalent diblock copolymer
problem realized here by percolation paths will be on hexagonal, rather than
square lattice. On the basis of universality we expect our results to extend 
also to the square lattice case.

\begin{figure}  
\centerline{
\psfig{file=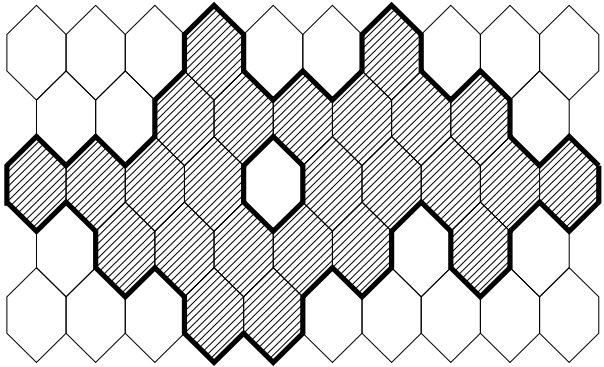,width=8.cm}}
\vskip 0.2truecm
\caption{Percolation cluster of connected occupied hexagons (dashed). Each 
hexagon is centered on a site of the dual, triangular lattice.}
\label{FIG12}
\end{figure}

Let us consider a percolation cluster as sketched in Fig. \ref{FIG12}. 
Its external
perimeter is a self-avoiding ring. The ensemble of all possible conformations
of an external hull on the lattice can be regarded as a problem of ring polymer
(grand canonical) statistics, as discussed in Refs. \cite{percolation}.
One further realizes that this effective ring polymer problem is characterized
by attractive interactions. These originate from the fact that, at threshold
($p = p_c = 1/2$), multiple visitations \cite{nota} by the hull of the same
occupied, or vacant, hexagon, give higher probability to the realization of a
ring configuration. Indeed, when the contour proceeds essentially straight,
to each new step corresponds a new hexagon whose state (occupied or vacant)  
has to be determined. So, each step implies a factor $p_c=1/2$ in the 
probability weight of the whole configuration.
When the contour folds on itself and revisits, after some steps, the perimeter
of the same hexagon, the factor $1/2$ does not apply, resulting in higher
global probability. This is equivalent to an attractive interaction favoring
the multiple visitations of the same hexagon.

It is convenient here to summarize some very recent exact results concerning 
the fractal dimensions of various percolative sets.
Following Ref. \cite{ADA99} we consider an annular region of the hexagonal
lattice delimited by an inner circle of small radius $r$, and an external one, 
of radius $R \gg r$. Two types of paths connecting the two circles are also
considered. These paths are formed by connected and self-avoiding sequences 
of either occupied, or empty hexagons.
The so-called {\it path-crossing} probability, namely the probability that $l$ 
non-overlapping paths connect inner to outer circles, was found to behave 
asymptotically as:
\begin{equation}
P_l (r,R) \approx (r/R)^{x_l}
\label{crossing}
\end{equation}
where:              
\begin{equation}
x_l = \frac{l^2 -1}{12}
\label{xl}
\end{equation}
The formula is valid if there is at least a path of each type, and the 
probability depends only on the total number of these paths, not on their 
type \cite{ADA99}.

\begin{figure}  
\centerline{
\psfig{file=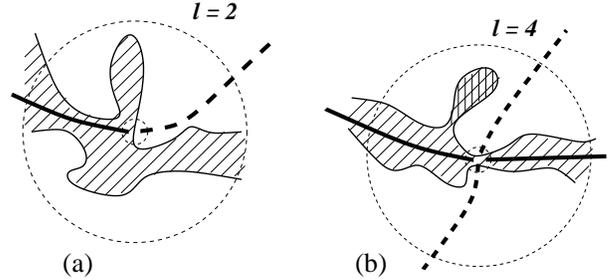,width=8.cm}}
\vskip 0.2truecm
\caption{Path crossing configurations for (a) a dashed and a solid line and 
(b) two dashed and two solid lines. The probabilities of the configurations 
yield the fractal dimensions of
the external perimeter of the hull (a) and of the cutting hexagons of the
backbone (b). The dashed region indicates the percolating cluster of occupied 
hexagons, while the double dashed region of (b) shows a dangling end, a part 
of the cluster which does not belong to the backbone.}
\label{FIG13}
\end{figure}

Figure \ref{FIG13} shows examples of crossing paths; without resolving the 
underlying lattice structure, we draw them as solid lines if they connect 
filled hexagons, while dashed lines are used for paths connecting empty 
hexagons.
As a first example (see Fig. \ref{FIG13}(a)) we consider the crossing 
probability for a continuous and a dashed path, which according to 
Eqs. (\ref{crossing},\ref{xl}) decays as $P_{l=2} \approx (r/R)^ {1/4}$.
One recognizes immediately that the set of points for which two of such 
self-avoiding paths can be drawn are those of the external perimeter or hull 
of the percolative cluster (see Fig. \ref{FIG13}(a)). This identification
allows to derive the fractal dimension of the hull. Since the area 
enclosed by the annulus is proportional to $R^2$, the perimeter of 
the hull enclosed in the annulus must scale as $L_{eh} \sim R^2 P_{l=2} \sim
R^{2-x_2} = R^{7/4}$. 
Identifying the external hull as a polymer ring at the $\Theta$-point one then 
derives that the latter has a fractal dimension $D_\Theta = D_{l=2} \equiv 
2 - x_{l=2} = 7/4$.

\begin{figure}  
\centerline{
\psfig{file=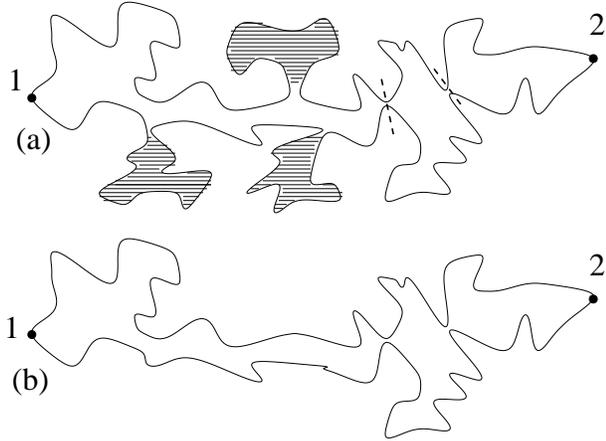,width=8.cm}}
\vskip 0.2truecm
\caption{(a) Schematic representation of a percolation cluster with dangling 
ends (dashed areas). When these are eliminated one remains with the cluster 
backbone (b). The dashed segments cut the cluster in correspondence to the 
so-called "red" hexagons.
}
\label{FIG14}
\end{figure}

In order to make contact with the diblock copolymer zipping, let us now
imagine to identify two points, 1 and 2, dividing the cluster hull in two
equally long parts (see Fig. \ref{FIG14}).
By fixing these two points on the cluster perimeter, one automatically defines
a backbone as a subset of the whole cluster. The backbone is the union of all
connected paths of occupied hexagons, which are strictly self-avoiding (i.e.
in each path a given hexagon appears at most once) and join the points 1 and 2.
In force of the definition, the backbone does not include the so called
"dangling ends", i.e. those branches of the cluster connected to the main
body by narrow bridges (i.e. by regions in which only one occupied hexagon
is available, making impossible for a self-avoiding path of hexagons to
penetrate and exit at the same time). An example of dangling end is also
schematically shown as double dashed area in Fig. \ref{FIG13}(b).

The two points we fix on the contour clearly divide into two sides the
perimeter of the backbone. Even if in this case it is not possible to give a
simple expression of the effective interactions determining the shape of the
two backbone sides, we expect them to be essentially local, like in the case
of the hull, and to act differently according to whether they involve close
encounters of the same side, or between different sides. This is consistent 
with the idea that the two sides of the backbone perimeter could
represent the statistics of a ring version of the diblock copolymer at the 
transition, the two parts corresponding, respectively, to blocks A and B.

For the calculation of the fractal dimension of the external perimeter of
the backbone one can use Eqs. (\ref{crossing},\ref{xl}) taking
two continuous paths and a dashed one.
This configuration clearly identifies the perimeter of the backbone.
Indeed, the two continuous paths guarantee that occupied hexagons
inside the interior circle belong to a whole path connecting two   
infinitely distant points. At the same time, a dashed path implies that
the vacant hexagons facing the occupied ones belong to the exterior
of the cluster, and thus, also of its backbone.

Therefore we now take $l=3$ for the exponents defined in Eq. (\ref{xl}). In 
this case we find that the external perimeter of the backbone scales as 
$L_{\rm bb} \sim R^{2-x_3} = R^{4/3}$, which implies a fractal dimension
$D_3 = 4/3$. This dimension is consistent with that found for the diblock 
copolymer at the zipping transition.
Furthermore, it is natural to associate the switching on of effective
attractive interactions between the two backbone sides to the existence of
narrow bottlenecks in the backbone itself (corresponding to only one hexagon).
These are the so-called cutting or "red" hexagons of the backbone
\cite{percobook}, which are visited by the two blocks simultaneously. 
In oder to determine their fractal dimension
one has to consider a percolative configuration with two continuous and 
two dashed paths joining the circles, as sketched in Fig. \ref{FIG13}(b). 
These identify a dimension $D_4=2 - x_4 = 3/4$.
Thus, for a backbone with external perimeter equal to $N$ and an average of
$N_{AB}$ contacts we find $N \sim R^{D_3}$ and $N_{AB} \sim R^{D_4}$.
Consequently, the average number of contacts between the backbone sides grows 
like $N_{AB} \sim N^{D_4/D_3} = N^{9/16}$.
By identifying the external perimeter of the backbone with the ring diblock 
copolymer at its transition, one eventually finds $\nu_c=3/4$ and $\phi=9/16$.
Now the numerical determinations of $\nu_c$ and $\phi$ at the zipping 
transition are remarkably consistent with these values, making the conjecture 
extremely plausible \cite{nota2}.
                        
\section{Conclusions}
\label{sec:conclusions}

In this paper we studied the phase diagram for the collapse transition of a 
diblock copolymer with attractive interactions between monomers of different 
species and a triple contact interaction $\delta$, which according to its sign 
may either favor, or unfavor, compactification.
In the region of negative $\delta$ we find a first order transition from a
swollen to a compact, spiral phase, while in the positive $\delta$ region
there is a sequence of a continuous zipping transition and a collapse of first 
order type to compact conformations at a lower temperature.

Our exact enumerations and Monte Carlo simulations yield numerical estimates 
of the critical exponents $\nu_c$ and $\phi$ of the zipping transition, which 
are consistent with those we could conjecture using recent results for the 
fractal dimensions of the percolation cluster backbone, from which we expect 
$\nu_c=3/4$ and $\phi =9/16$. 
The numerically determined exponents, therefore, are strongly supporting
the hypothesis that the transition admits a description in terms of percolative
stochastic geometry: the two blocks of the copolymer have the same fractal 
geometry as the two sides of a cluster backbone, and their contacts correspond 
to the cutting hexagons, or links of the same backbone. 
This is, to our knowledge, the second example of a percolative representation 
for a polymer conformational transition in 2D, besides that of the 
$\Theta$-point. The common percolative roots 
of these transitions suggests the possibility of a deep link between them,
which ought to be elucidated by further studies.

The results obtained for the various transitions appearing in the phase 
diagram help in clarifying the nature of the adsorption-like collapse occurring 
at $\delta=0$ and first detected in Ref. \cite{OSS00}. In spite of the fact
that most tests are not able to put into clear evidence the existence of two 
successive transitions, the only multicritical behavior which can be 
characterized coming from the
high temperature region seems to definitely belong to the universality class of
the continuous zipping transition identified for positive values of $\delta$.
Besides the compatibility of the exponent estimates, a very strong support to
such conclusion comes from our determination of the universal amplitude ratio
between squared end-to-end distance and radius of gyration of the polymer.

Another interesting aspect of the phase diagram calculated in this paper are
the first order swollen-collapsed and zipped-collapsed transitions found,
respectively, for negative and positive values of $\delta$.
In particular the latter resembles the transition from swollen to spiral
state found in {\it oriented} polymers \cite{oriented}, i.e. chains to which 
an overall orientation is assigned and where different energies are associated 
to contacts between parallel or antiparallel segments 
of the chain. In fact the analogy between the diblock copolymer in a zipped
state and an oriented polymer is very appropriate: in the zipped diblock
parallel contacts are of AB type, while antiparallel ones are 
contacts between monomers of equal type. Different energies are clearly
associated with the two types of contacts.

It is worthwhile to recall that simple polymer models with some sort of 
zipping transition attracted already some attention in the recent literature 
\cite{Imbert,Causo}, mainly because of the relevance that such transition can 
have for biopolymers. 
Imbert {\it et al.} \cite{Imbert} considered a diblock formed by two strands
of oppositely charged monomers interacting with each other through long range 
Coulomb forces and found evidence of the existence of a zipping transition
followed by a collapse at lower temperatures. 
Causo {\it et al.} \cite{Causo} considered a simple model for the DNA
denaturation transition, in which only the monomers which are at equal 
distances along the sequence from the center of the chain interact. They 
found evidence of a first order transition, from a swollen to a zipped phase. 
By its construction their model has no other transitions to compact state.
In their case the first order zipping seems to be due to the selective 
interactions of monomers along the chain.
Also in our model, if we turn on interactions only between AB monomers at equal 
distance from the center we find evidence of a first order zipping transition.

Finally, we point out that there are several possible extensions of 
this work. First of all, it would be interesting to generalize the model to 
three dimensions and to investigate the properties of the zipping transition in 
that case \cite{OSS00}.
Another open issue is the effect of disorder on the interaction between 
monomers for the zipping transition, which would allow to understand the 
behavior of models of polymers more relevant for applications to chemistry
or biology than a simple diblock.

We thank F. Seno for discussions and collaboration in the early stages 
of this work. Financial support by MURST through COFIN 1999 and INFM through 
PAIS 1999 is gratefully acknowledged.

\end{multicols}


\begin{references}
\bibitem{nota} Visitation means here that the hull reaches the hexagon 
perimeter, or returns to it after having left.
\bibitem{nota2}  Note that in Ref. \cite{OSS00} an identification with the two 
halves of the externally accessible hull, rather than the backbone perimeter
was proposed. Although this geometrical contour has also a fractal dimension
$4/3$, problems then arise in the discussion of the contacts of these two 
parts, which are expected to have fractal dimension zero rather than $3/4$.
The present argument solves this problem and is fully consistent.
\end{references}
\end{document}